\documentclass[%
 reprint,
superscriptaddress,
amsmath,amssymb,
aps,
prc,
floatfix,
]{revtex4-2}

\usepackage{graphicx}
\usepackage{dcolumn}
\usepackage{bm}
\usepackage{subcaption}
\usepackage{xcolor}
\usepackage{natbib}
\newcommand{\nuc}[2]{$^{#1#2}$}
\newcommand{\g}{$\gamma$}

\DeclareCaptionJustification{justified}{\leftskip=0pt \rightskip=0pt \parfillskip=0pt plus 1fil}
\begin{document}
\preprint{APS/123-QED}

\title{Cross Section Measurement of the \nuc82Kr(p,\g)\nuc83Rb Reaction in Inverse Kinematics}
% Authors
\author{A. Tsantiri}\email{tsantiri@frib.msu.edu}
\affiliation{Department of Physics and Astronomy, Michigan State University, East Lansing, MI 48824, USA}
\affiliation{Facility for Rare Isotope Beams, Michigan State University, East Lansing, MI 48824, USA}
\affiliation{Joint Institute for Nuclear Astrophysics - Center for the Evolution of the Elements, East Lansing, MI 48824, USA}

\author{A. Palmisano-Kyle}\altaffiliation[Present Address:]{Physics Department, University of Tennessee, TN 37996, USA}
\affiliation{Department of Physics and Astronomy, Michigan State University, East Lansing, MI 48824, USA}
\affiliation{Facility for Rare Isotope Beams, Michigan State University, East Lansing, MI 48824, USA}
\affiliation{Joint Institute for Nuclear Astrophysics - Center for the Evolution of the Elements, East Lansing, MI 48824, USA}

\author{A. Spyrou}
\affiliation{Department of Physics and Astronomy, Michigan State University, East Lansing, MI 48824, USA}
\affiliation{Facility for Rare Isotope Beams, Michigan State University, East Lansing, MI 48824, USA}
\affiliation{Joint Institute for Nuclear Astrophysics - Center for the Evolution of the Elements, East Lansing, MI 48824, USA}

\author{P. Mohr}
\affiliation{Institute for Nuclear Research (Atomki), H-4001 Debrecen, Hungary}

\author{H. C. Berg}
\affiliation{Department of Physics and Astronomy, Michigan State University, East Lansing, MI 48824, USA}
\affiliation{Facility for Rare Isotope Beams, Michigan State University, East Lansing, MI 48824, USA}
\affiliation{Joint Institute for Nuclear Astrophysics - Center for the Evolution of the Elements, East Lansing, MI 48824, USA}

\author{P. A. DeYoung}
\affiliation{Physics Department, Hope College, Holland, MI 49423, USA}

\author{A. C. Dombos}
\altaffiliation{Present Address: Department of Physics, University of Notre Dame, IN 46556, USA}
\affiliation{Department of Physics and Astronomy, Michigan State University, East Lansing, MI 48824, USA}
\affiliation{Facility for Rare Isotope Beams, Michigan State University, East Lansing, MI 48824, USA}
\affiliation{Joint Institute for Nuclear Astrophysics - Center for the Evolution of the Elements, East Lansing, MI 48824, USA}

\author{P. Gastis}
\altaffiliation{Present Address: Los Alamos National Laboratory, New Mexico 87545, USA}
\affiliation{Department of Physics, Central Michigan University, Mt Pleasant, MI 48859, USA}
\affiliation{Joint Institute for Nuclear Astrophysics - Center for the Evolution of the Elements, East Lansing, MI 48824, USA}

\author{E. C. Good}
\affiliation{Facility for Rare Isotope Beams, Michigan State University, East Lansing, MI 48824, USA}
\affiliation{Joint Institute for Nuclear Astrophysics - Center for the Evolution of the Elements, East Lansing, MI 48824, USA}

\author{C. M. Harris}
\affiliation{Department of Physics and Astronomy, Michigan State University, East Lansing, MI 48824, USA}
\affiliation{Facility for Rare Isotope Beams, Michigan State University, East Lansing, MI 48824, USA}
\affiliation{Joint Institute for Nuclear Astrophysics - Center for the Evolution of the Elements, East Lansing, MI 48824, USA}

\author{S. N. Liddick}
\affiliation{Department of Chemistry, Michigan State University, East Lansing, MI 48824, USA}
\affiliation{Facility for Rare Isotope Beams, Michigan State University, East Lansing, MI 48824, USA}
\affiliation{Joint Institute for Nuclear Astrophysics - Center for the Evolution of the Elements, East Lansing, MI 48824, USA}

\author{S. M. Lyons} 
\altaffiliation{Present Address: Pacific Northwest National Laboratory, WA 99354, USA}
\affiliation{Department of Physics and Astronomy, Michigan State University, East Lansing, MI 48824, USA}
\affiliation{Facility for Rare Isotope Beams, Michigan State University, East Lansing, MI 48824, USA}
\affiliation{Joint Institute for Nuclear Astrophysics - Center for the Evolution of the Elements, East Lansing, MI 48824, USA}

\author{O. Olivas-Gomez}\altaffiliation{Present Address: Johns Hopkins University Applied Physics Laboratory, MD, 20723, USA}
\affiliation{Department of Physics, University of Notre Dame, Notre Dame, IN 46556, USA}

\author{G. Owens-Fryar}
\affiliation{Department of Physics and Astronomy, Michigan State University, East Lansing, MI 48824, USA}
\affiliation{Facility for Rare Isotope Beams, Michigan State University, East Lansing, MI 48824, USA}
\affiliation{Joint Institute for Nuclear Astrophysics - Center for the Evolution of the Elements, East Lansing, MI 48824, USA}

\author{J. Pereira} 
\affiliation{Department of Physics and Astronomy, Michigan State University, East Lansing, MI 48824, USA}
\affiliation{Facility for Rare Isotope Beams, Michigan State University, East Lansing, MI 48824, USA}
\affiliation{Joint Institute for Nuclear Astrophysics - Center for the Evolution of the Elements, East Lansing, MI 48824, USA}

\author{A. L. Richard}
\altaffiliation{Present Address: Lawrence Livermore National Laboratory, CA 94550, USA}
\affiliation{Department of Physics and Astronomy, Michigan State University, East Lansing, MI 48824, USA}
\affiliation{Facility for Rare Isotope Beams, Michigan State University, East Lansing, MI 48824, USA}
\affiliation{Joint Institute for Nuclear Astrophysics - Center for the Evolution of the Elements, East Lansing, MI 48824, USA}

\author{A. Simon} 
\affiliation{Department of Physics, University of Notre Dame, Notre Dame, IN 46556, USA}

\author{M. K. Smith}
\affiliation{Department of Physics and Astronomy, Michigan State University, East Lansing, MI 48824, USA}
\affiliation{Facility for Rare Isotope Beams, Michigan State University, East Lansing, MI 48824, USA}

\author{R. G. T. Zegers}
\affiliation{Department of Physics and Astronomy, Michigan State University, East Lansing, MI 48824, USA}
\affiliation{Facility for Rare Isotope Beams, Michigan State University, East Lansing, MI 48824, USA}
\affiliation{Joint Institute for Nuclear Astrophysics - Center for the Evolution of the Elements, East Lansing, MI 48824, USA}

\begin{abstract}
The total cross section of the \nuc82Kr(p,\g)\nuc83Rb reaction was measured for the first time at effective center-of-mass energies between 2.4 and 3.0 MeV, within the relevant Gamow window for the astrophysical \g\, process. The experiment took place at the National Superconducting Cyclotron Laboratory at Michigan State University using the ReA facility. A \nuc82Kr beam was directed onto a hydrogen gas cell located at the center of the Summing NaI(Tl) (SuN) detector. The obtained spectra were analyzed using the \g-summing technique and the extracted cross section was compared to standard statistical model calculations using the \textsc{non-smoker} and \textsc{talys} codes. The comparison indicates that standard statistical model calculations tend to overproduce the cross section of the \nuc82Kr(p,\g)\nuc83Rb reaction relative to the experimentally measured values. Furthermore, the experimental data was used to provide additional constraints on the nuclear level density and \g-ray strength function used in the statistical model calculations.
\end{abstract}
\maketitle

\section{\label{sec:intro}Introduction}
\par One of the most fundamental questions in nuclear astrophysics relates to understanding the mechanisms through which the elements are forged in the stars. For the vast majority of the elements heavier than iron, stellar nucleosynthesis is largely governed by the slow \textit{s}- and rapid \textit{r}- neutron capture processes \cite{b2fh,r-review,s-review}, as well as possible contributions from the intermediate \textit{i} process \cite{i-process_2016,i-process_1977}. % and \textit{n}-process \cite{n-process}. 
However, a relatively small group of naturally occurring, neutron-deficient isotopes, located in the region between \nuc74Se and \nuc19$^6$Hg, the so called \textit{p} nuclei, cannot be formed by the neutron capture processes \cite{Rauscher_2016}. These $\approx$ 30 stable nuclei are believed to be formed in the commonly called \g\, process from the ``burning" of preexisting \textit{r}- and \textit{s}-process seeds in stellar environments of sufficiently high temperatures of $2 \text{ GK} \leq T \leq 3.5$ GK, where a sequence of photodisintegration reactions can occur \cite{Rauscher_2013}. The astrophysical site where such temperature conditions are fulfilled has been a subject of controversy for more than 60 years and is currently believed to occur in the ONe layers of Type II supernovae \cite{Prantzos_1990,Woosley_1978,Travaglio_2011}, and/or in thermonuclear Type Ia supernovae \cite{Nishimura_2017}.
\par In order to reproduce the \textit{p}-nuclei abundances that are observed in nature, networks of nuclear reactions are simulated under appropriate astrophysical conditions. However, in addition to any astrophysical uncertainties, many nuclear uncertainties enter these network calculations, since there are almost 20 000 nuclear reactions on almost 2 000 nuclei that need to be taken into account \cite{Arnould_2003}. The nuclear physics inputs required for these calculations consists mainly of reaction rates that need to be experimentally constrained \cite{rapp06}.
For the case of the \g~ process, the dominant reactions are photodisintegration reactions. As mentioned in Ref.~\cite{Rauscher_book} it is often advantageous to measure the exothermic reverse reaction. This is because in an astrophysical environment, the ground state contribution to the photodisintegration reaction rate can be small compared to the full reaction rate. 
It should be noted that the majority of nuclei involved in the \g\, process are radioactive, and experimental data are almost non-existent. Therefore, the associated uncertainties in the predicted reaction rates tend to increase significantly when moving away from stability \cite{rapp06}. 
\par Despite the decades of considerable experimental effort \cite{cologne2022,Lotay_2021,Cheng_2021,Simon2019, Foteinou_Se_2018,atomki2010,Spyrou_2007}, experimental cross sections of \g-process reactions are mostly unknown and the related reaction rates are based primarily on Hauser-Feshbach (HF) theoretical calculations \cite{HF}. In calculating reaction rates through HF theory, nuclear properties such as nuclear level densities (NLDs) and \g-ray strength functions (\g SFs) are used as input.
Constraining nuclear input in the HF model remains a challenge and it is therefore crucial to provide new experimental cross sections relevant to the \g\, process. The present work contributes to this larger effort to constrain reaction theory by reporting on the first measurement of the \nuc82Kr(p,\g)\nuc83Rb reaction cross section. The measurement took place at effective center-of-mass energies between 2.4 and 3.0 MeV, which are within the relevant Gamow window for the \g\, process which lies between 1.9 and 4.0 MeV.

\section{\label{sec:experiment}Experimental Details}
The present measurements were carried out at the National Superconducting Cyclotron Laboratory at Michigan State University using the ReA reaccelerator facility \cite{ReA} to accelerate a stable \nuc82Kr$^{27+}$ beam at energies of 3.1, 3.4 and 3.7 MeV/nucleon. The delivered beam impinged on a hydrogen gas-cell target and the \g\, rays produced by the reaction were detected by the Summing NaI(Tl) (SuN) detector. Details on the experimental setup are provided in Ref.~\cite{Alicia}, but are briefly summarized here.
\par The hydrogen gas-cell target, located in the center of the SuN detector, was made of plastic with 2-$\mu$m thick molybdenum foils used as entrance and exit windows. The cell had a length of 4 cm and included a tantalum ring on the upstream side and tantalum foil lining the inner walls of the cell, to shield the plastic from the beam and reduce beam-induced background. The hydrogen gas inside the cell was kept at a pressure of $\approx$ 600 Torr.
\par The SuN detector is a 4$\pi$ calorimeter with the shape of a $16\times16$ inch barrel with a 1.8 inch diameter borehole along its axis. The barrel is segmented into 8 optically isolated NaI(Tl) crystals, each connected to three photomultiplier tubes (PMTs). A detailed description of the SuN detector and its data acquisition system can be found in Ref.~\cite{SuN_paper2013}. By positioning the target at the center of SuN, the large angular coverage and high detection efficiency of the detector allowed for the application of the \g-summing technique \cite{Spyrou_2007,SuN_paper2013}. In this way, the spectra obtained by the individual segments provide sensitivity to the individual \g-ray transitions, whereas the full energy deposited in SuN provides sensitivity to the populated excitation energies. For this reason, three main spectra are used in SuN data analysis: sum of segments (SoS), total absorption spectra (TAS), and multiplicity. SoS corresponds to the energy detected in the individual segments, TAS to the full energy deposited in the detector, and multiplicity indicates how many segments of SuN recorded energy in each event. 
\par In order to reduce cosmic-ray induced background and increase the sensitivity of the SuN detector, the Scintillating Cosmic Ray Eliminating ENsemble (SuNSCREEN) was positioned above SuN and was utilized as a veto detector \cite{SuNSCREEN}. SuNSCREEN is a plastic scintillator detector array comprised of nine bars, each with two PMTs, forming a roof-like arrangement above the SuN detector. To reduce the cosmic-ray induced background, a veto gate was applied to all events that recorded signals in both PMTs of a SuNSCREEN bar, and at least one segment of SuN.

\section{\label{sec:analysis}Analysis}
\par The reaction cross section, $\sigma$, can be calculated as
\begin{equation}\label{cs}
    \sigma = \frac{ Y }{N_b\, N_t\, \epsilon}
\end{equation}
where $Y$ is the experimental yield, namely how many reactions of interest were measured, $N_b$ is the number of projectiles, $N_t$ is the areal target density, and $\epsilon$ is the detection efficiency. The number of projectiles was calculated from the current measured off of the beam pipe which was used as a Faraday cup, taking into account the beam charge state of 27$^+$. The areal target density was calculated based on the size of the gas cell and the average recorded gas pressure during each measurement. 
\par In order to avoid the assumption that all reactions take place at the center of the gas cell, the effective center-of-mass energy, $E_{eff}$, was calculated. $E_{eff}$ corresponds to the beam energy in the target at which one-half of the yield for the full target thickness is obtained \cite{cauldrons}. Assuming a linear decrease in cross section from the entrance of the target to the exit, the effective energy is calculated as:
\begin{equation}
    E_{eff} = E_0 - \Delta E +\Delta E \left\{-\frac{\sigma_2}{\sigma_1-\sigma_2}+\left[\frac{\sigma_1^2+\sigma_2^2}{2(\sigma_1-\sigma_2)^2}\right]^{1/2}\right\}
 \end{equation}
 where $\sigma_1$ is the cross section at the entrance of the target for incident beam energy $E_0$, $\Delta E$ is the energy loss within the target, and $\sigma_2$ is the cross section at the exit of the target at $E_0 - \Delta E$. The values for $\sigma_1$ and $\sigma_2$ were obtained from \textsc{non-smoker} \cite{nonsmoker}. The ratio of $\sigma_1/\sigma_2$ was 1.6, 1.7 and 2.0 for initial beam energy 3.7, 3.4 and 3.1 MeV/nucleon respectively. The resulting $E_{eff}$ was in agreement within error with the center-of-mass energy at the center of the gas cell $E_0 - \Delta/2$.    
 
\par In the following section, more details regarding the calculation of the ratio of the yield and the efficiency will be discussed.

\subsection{\label{g-summing}The \g-summing technique}
When a proton from the gas-target is captured by the \nuc82Kr beam, it populates an excited state of \nuc83Rb of energy $E_x = E_{cm} + Q$, where $E_{cm}$ is the beam energy in the center-of-mass system and $Q = 5.77$ MeV is the \nuc82Kr(p,\g)\nuc83Rb reaction $Q$ value. For the present experiment the excitation energy of \nuc83Rb ranges from 8 to 9 MeV. The \nuc83Rb compound nucleus can de-excite through many different possible \g-ray cascades. The emitted \g\, rays are recorded by SuN. By adding the energy of all \g\, rays originating from a single cascade, a ``sum peak" is produced in the total \g-summed spectrum with energy equal to $E_x$ \cite{SuN_paper2013}. The integral of the sum peak corresponds to the experimental yield, $Y$. The efficiency of the sum peak depends not only on the energy $E_x$, but also on the average \g-ray multiplicity of the cascade as discussed in Section~\ref{efficiency_yield}. Due to the experiment being conducted in inverse kinematics with a gas-target, the sum peak has a larger energy range than in forward kinematics measurements due to Doppler shift and energy straggling through the molybdenum foil and the hydrogen gas. For this reason a recently developed analysis technique was applied to this measurement. The technique was first demonstrated for the \nuc84Kr(p,\g)\nuc85Rb cross section in Ref.~\cite{Alicia}.

\subsection{\label{background_sub}Background subtraction}
The background in this experiment can be attributed to two major contributors: cosmic-ray induced or room background and beam-induced background. The contribution of the cosmic-ray induced background is reduced with two ways: the SuNSCREEN veto and beam pulsing. More specifically, the events that were recorded by the two PMTs of one of SuNSCREEN's scintillator bars in coincidence with a segment of SuN were rejected, as mentioned in Sec.~\ref{sec:experiment}. In order to remove any background events that were not accounted for through the SuNSCREEN veto, the \nuc82Kr beam was pulsed using the EBIT charge buncher \cite{ReA}. The beam was delivered in 100 $\mu$s pulses separated by 200 ms of dead time. 
Two data sets of 100 $\mu$s were recorded for each beam pulse, one while beam was delivered, and one during the dead time to record room background that was subtracted from the final spectra.
\par In order to account for beam-induced background, data was acquired with the cell full of hydrogen gas, and with the cell empty. The empty-cell data were normalized to the beam current, and subtracted from the full-cell data to obtain the final spectra.
\par
\begin{figure}[htbp!]
    \includegraphics[width=0.48\textwidth]{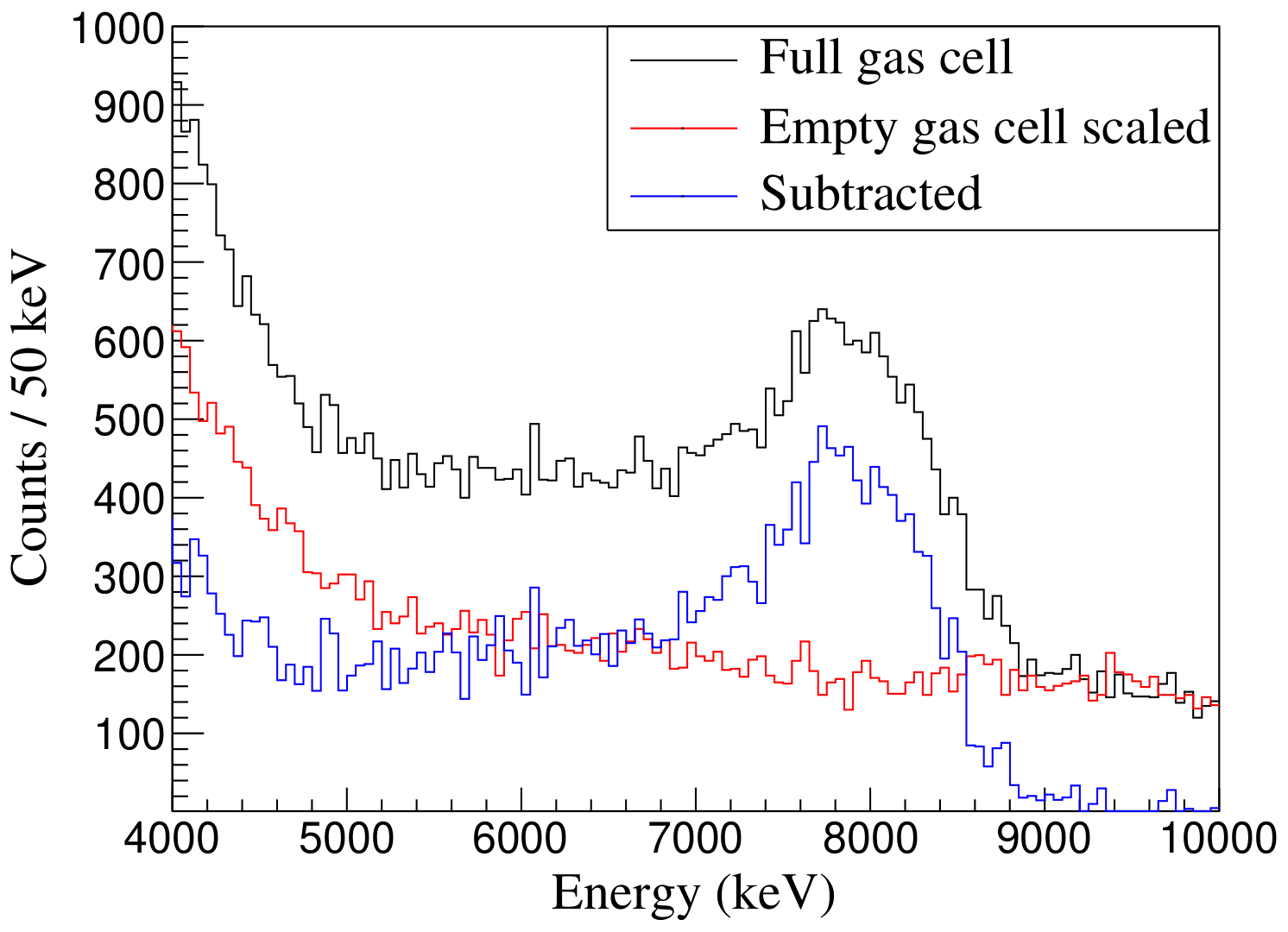}
    \caption{Doppler corrected TAS spectra showing the background subtraction for the sum peak for the \nuc82Kr(p,\g)\nuc83Rb reaction at initial beam energy 3.7 MeV/nucleon. The black histogram corresponds to the gas cell filled with hydrogen gas, the red histogram to the empty gas cell scaled to the beam current, and the blue histogram is the fully subtracted sum peak that was used for the remaining analysis.}
    \label{fig:bgr_sub}
\end{figure}

 Doppler-shift corrections were applied on a segment-by-segment basis as described in Ref.~\cite{quinn_doppler}. 
Through this correction the \g-ray energy detected by each segment is reconstructed based on the average velocity of the recoil nucleus, as well as the different detection angles. 
The fully background subtracted and Doppler-shift corrected sum peak at initial beam energy 3.7 MeV/nucleon is shown in Fig.~\ref{fig:bgr_sub}.

\subsection{\label{efficiency_yield}Efficiency and Yield determination}
The efficiency of the SuN detector was dependent on both the energy of the individual \g\, rays, as well as the multiplicity of the detected cascades \cite{SuN_paper2013}. Furthermore, the large width of the sum peak (due to the Doppler effect and energy straggling through the gas cell target) required the detection efficiency of this measurement to be calculated as a function of the contribution of each possible excitation energy of the \nuc83Rb compound nucleus (CN). The $Q$ value of the \nuc82Kr(p,\g)\nuc83Rb reaction was sufficiently high for the excitation energy, $E_x$, of the CN to be in the nuclear continuum region where statistical model calculations are valid. Hence, the contribution of each possible $E_x$ of the CN can be simulated in order to extract the energies of the \g\, rays involved in the de-excitation of the CN based on SuN’s multiplicity, SoS, and TAS. This analysis technique was developed and validated for \nuc90Zr(p,\g)\nuc91Nb cross-section data and applied successfully for the measurement of the \nuc84Kr(p,\g)\nuc85Rb cross section in Ref.~\cite{Alicia}.

\par For the simulation of the \g\, deexcitation of the \nuc83Rb nucleus the \textsc{rainier} code \cite{rainier} was implemented. \textsc{Rainier} is a Monte Carlo code that simulates the de-excitation of a compound nucleus using statistical nuclear properties.
Within \textsc{rainier}, the user inputs the nuclear level structure of the nucleus under study. Namely, the low energy portion of the \nuc83Rb level scheme was taken from Ref.~\cite{ripl} up to 1.8 MeV, where the level scheme was considered to be complete. The upper portion was constructed using a combination of the NLD described through the Constant Temperature (CT) model \cite{ctm,ctm2} as well as the Back Shifted Fermi Gas (BSFG) model \cite{bsfg,bsfg2}. The user also inputs the $E_x$ and $J^\pi$ of the entry state, as well as the \g SF model parameters for the subsequent de-excitation. For the \g SF a Generalized Lorentzian of the form of Kopecky and Uhl \cite{kopecky-uhl} was adopted.

\par
\begin{figure}[htbp!]
    \begin{subfigure}[ht!]{0.48\textwidth}
        \includegraphics[width=\textwidth]{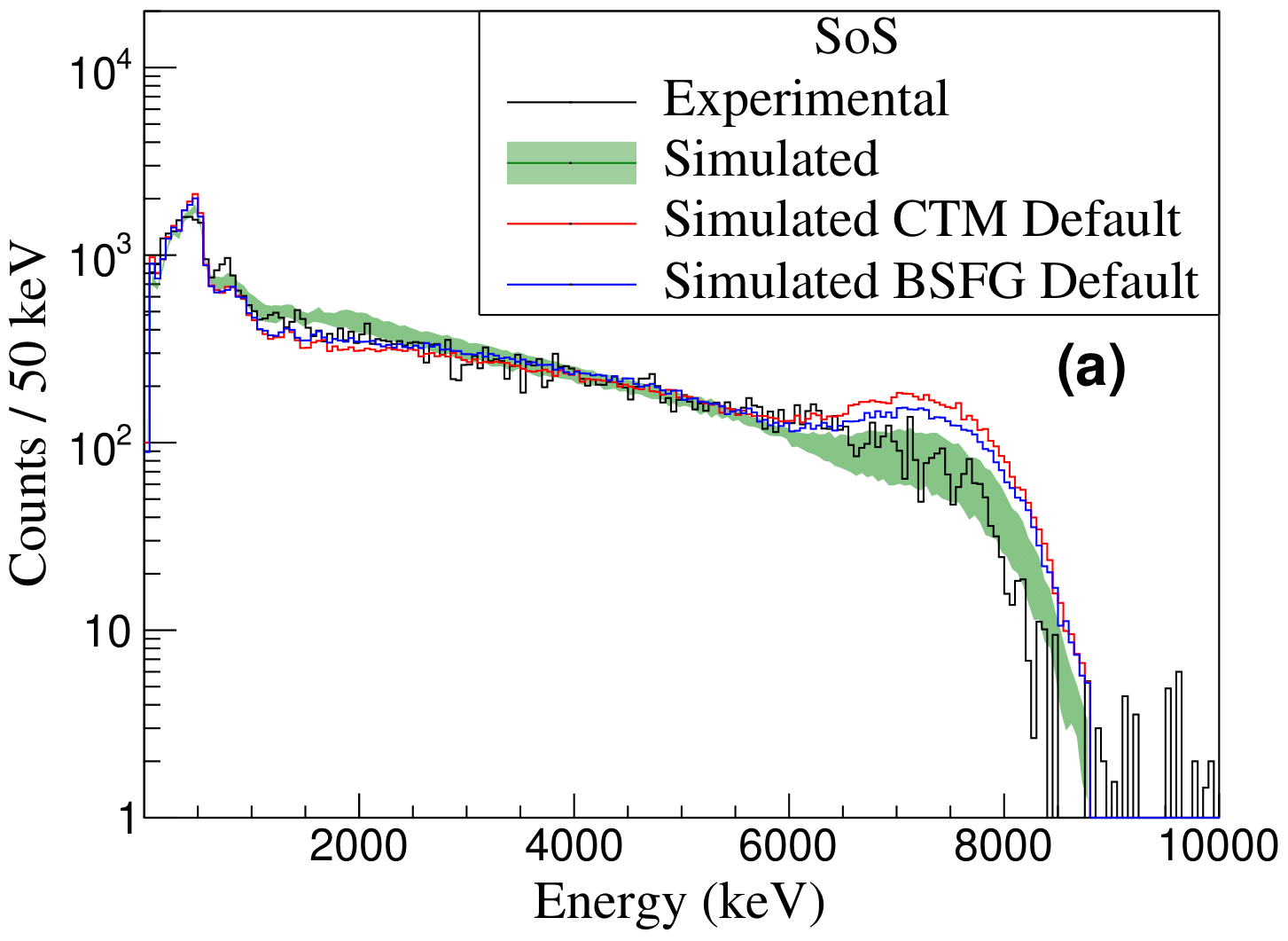}
    \end{subfigure}
    \begin{subfigure}[ht!]{0.48\textwidth}
        \includegraphics[width=\textwidth]{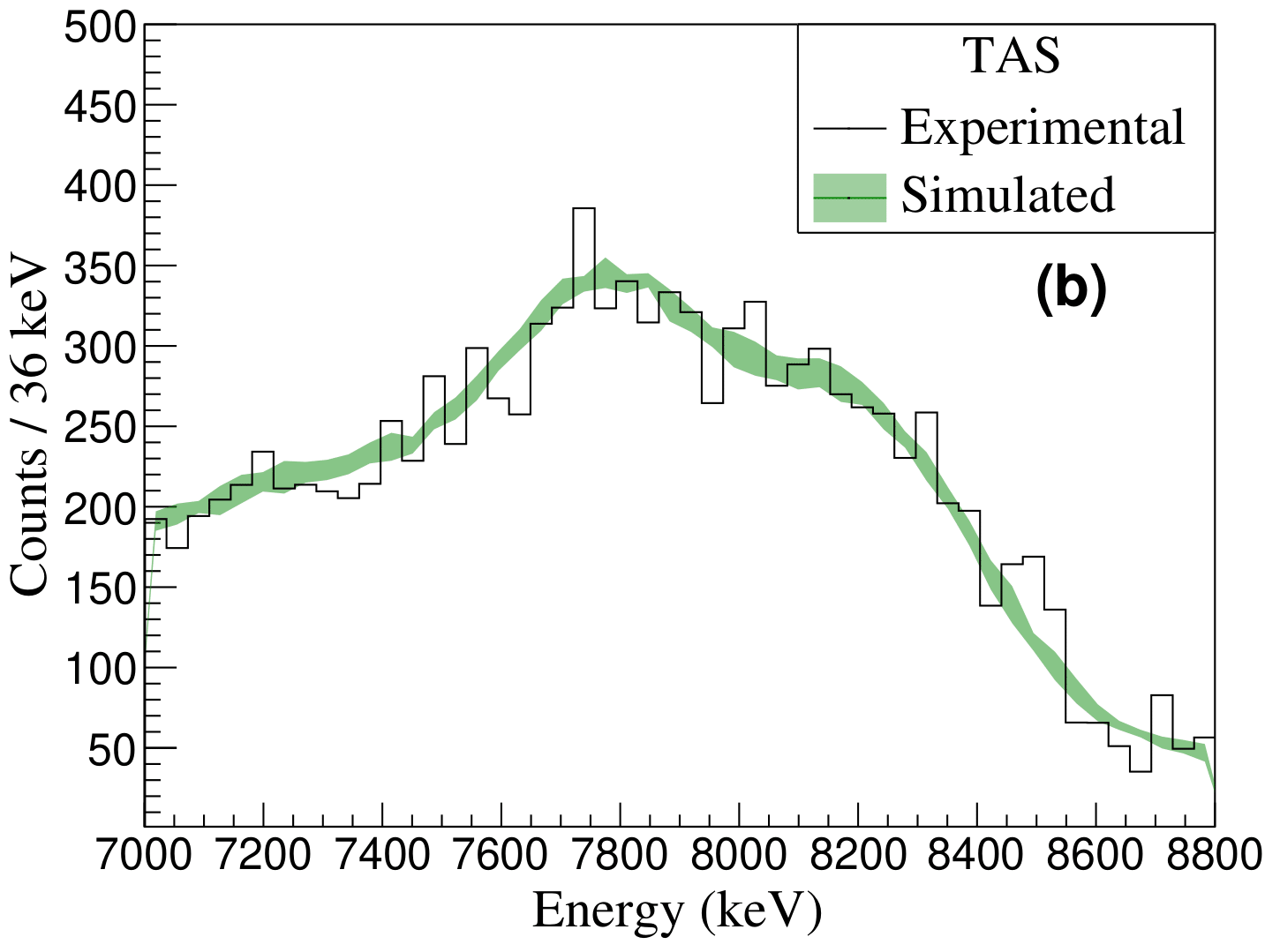}
    \end{subfigure}
    \begin{subfigure}[ht!]{0.48\textwidth}
        \includegraphics[width=\textwidth]{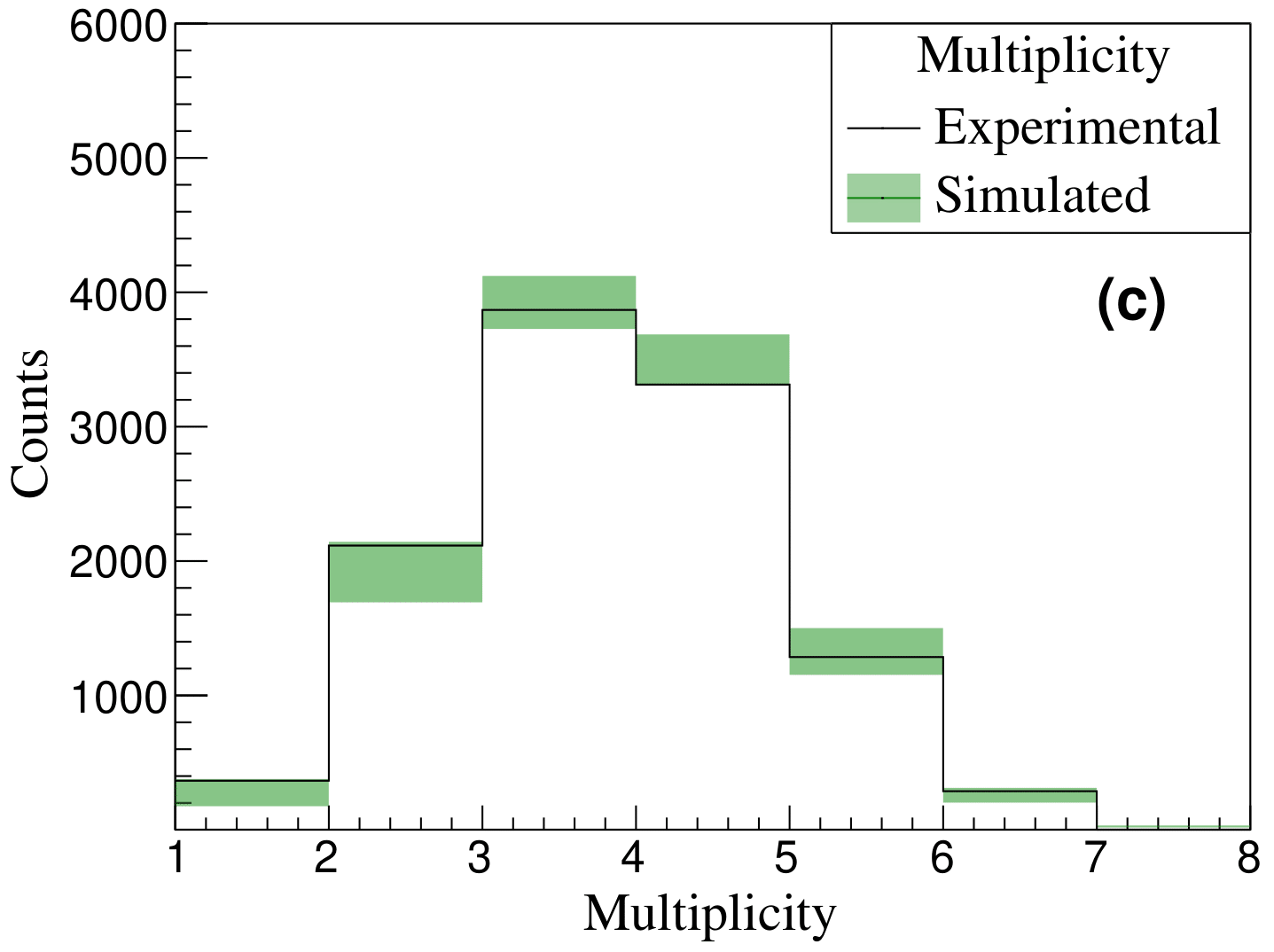}
    \end{subfigure}
    \caption{The $\chi^2$ minimization fits for the SoS (a), TAS (b) and multiplicity (c) for the \nuc82Kr(p,\g)\nuc83Rb reaction at an initial beam energy 3.7 MeV/nucleon. The black lines are the experimental spectra and the green bands indicate the simulated spectra for the combinations of the NLD and \g SF models chosen, taking into account all possible contributing $E_x$ of the \nuc83Rb CN. In (a) the red and blue lines are the simulated spectra for the default initial parameters of the BSFG and CT model NLD from Ref.~\cite{E&B2009} and \g SF from Ref.~\cite{talys_manual}. }
    \label{fig:chi2}
\end{figure}

 The choice of the NLD and \g SF model significantly affects the \g\, rays that can be emitted through the de-excitation of a nuclear level in the continuum. Therefore, the \g\, rays shown in a simulated SoS are highly dependent on the choice of the NLD and \g SF that are input in \textsc{rainier}. For this reason, the initial parameters of the NLD and \g SF models were varied, in order to replicate the experimental SoS spectra obtained by the de-excitation of the \nuc83Rb CN decay. The goal of this analysis was to identify suitable products of the NLD and \g SF, and not to constrain each one individually. The combinations that reproduced the SoS spectra are indicated by the green band shown in Fig.~\ref{fig:chi2}(a) along with the default parameters for \g SF and NLD through the CT and BSFG models. The default parameters for the NLD were obtained through Ref.~\cite{E&B2009}, and for the \g SF through Ref.~\cite{talys_manual}. Figure~\ref{fig:chi2}(a) indicates that the default model parameters fail to reproduce the experimental data, thus demonstrating the need to vary these parameters. It is noteworthy to mention that within these parameter combinations are the CT model parameters by Hoffman \textit{et al.} in Ref.~\cite{hoffman2004}, as well as a low-energy upbend on the M1 \g SF as parameterized by Guttormsen \textit{et al.} in Ref.~\cite{mo_upbend}. More detailed information on the choice of parameters is provided in Sec.~\ref{sec:theoretical}.

\par The \g\, rays obtained by the deexcitation of each contributing $E_x$ of the \nuc83Rb CN through \textsc{rainier} were then input in \textsc{Geant4} simulations \cite{geant}, in order to account for the  SuN detector's response function. To extract the overall contribution of each $E_x$ into the sum peak, a $\chi^2$ minimization code was utilized. The $\chi^2$ code uses the simulated TAS, SoS and multiplicity outputs of \textsc{Geant4} for each $E_x$, as well as the experimental spectra gated on the sum peak. Then it calculates the contribution of each $E_x$ required to replicate the shape of the experimental data by fitting the simulated SoS, TAS and multiplicity spectra simultaneously. The $\chi^2$ minimization output for the three spectra is shown in Fig.~\ref{fig:chi2}.

\par The ratio of the yield of the measurement, $Y$, over the detector's efficiency, $\epsilon$, corresponds to the total number of reactions that occurred. This ratio was obtained from the linear combination of the integrals of the simulated sum peak for each $E_x$, weighted based on each energy's contribution as extracted from the $\chi^2$ minimization output. The uncertainty of this ratio varied between 17\% and 23\% and is mostly attributed to the various parameters chosen for the NLD and \g SF models, as shown by the green bands in Fig.~\ref{fig:chi2}. 

\section{\label{sec:results}Results \& Discussion}
\begin{figure}[ht!]
    \centering
    \includegraphics[width=0.48\textwidth]{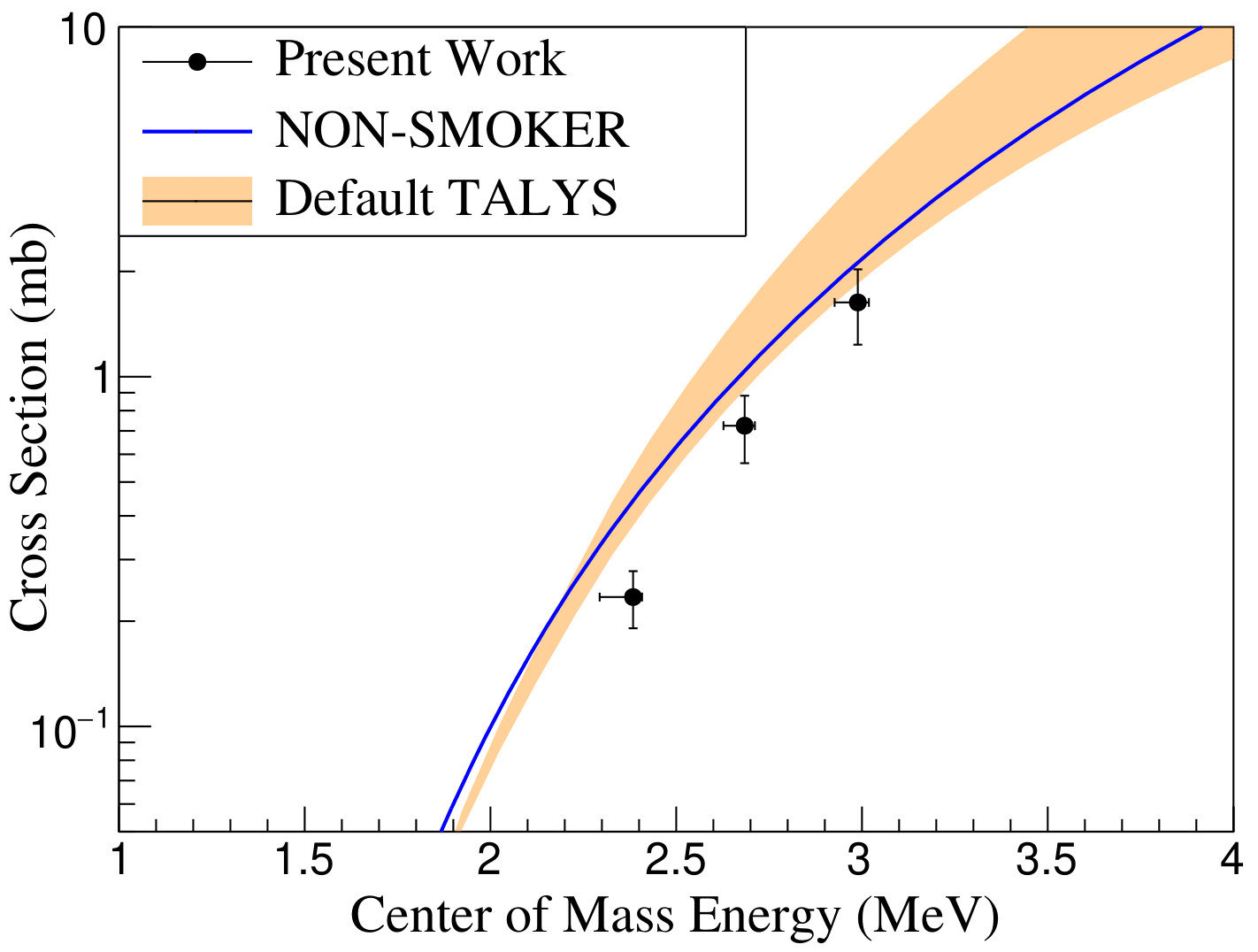}
    \caption{The measured cross section of the \nuc82Kr(p,\g)\nuc83Rb reaction (black dots) compared with standard \textsc{non-smoker}  theoretical calculations \cite{nonsmoker} (blue solid line) and default \textsc{talys} 1.96 calculations \cite{talys} (orange band). Standard statistical model calculations with the default models tend to overproduce the cross section relative to the experimentally measured values. A more consistent theoretical description of the experimental results is obtained in Section~\ref{sec:theoretical}.}
    \label{fig:cs}
\end{figure}

The cross section for the \nuc82Kr(p,\g)\nuc83Rb reaction as calculated using Eq.~(\ref{cs}) is presented in Table \ref{cs_table} and in Fig.~\ref{fig:cs}. The first column of Table \ref{cs_table} represents the initial beam energy that was impinged on the Mo foil and the second column the effective energy, $E_{eff}$. The third column shows the total number of incident beam particles on target and the fourth column represents our efficiency in detecting \g~rays from the de-excitation of the \nuc83Rb compound nucleus. As discussed in Section \ref{efficiency_yield}, the detection efficiency depends on the energy of the individual \g~rays, as well as the multiplicity of the cascade, and is extracted through simulations from the ratio of the yield over the efficiency.

\begin{table*}[ht!]
    \centering 
    \caption{Measured cross section of the \nuc82Kr(p,\g)\nuc83Rb reaction. The first column represents the initial beam energy in the lab system, the second column the center-of-mass effective energy of the reaction, and the third shows the total number of incident beam particles on target. The fourth column represents our efficiency in detecting \g-rays from the de-excitation of the \nuc83Rb compound nucleus (which depends on the energy of the individual \g rays, and the multiplicity of the cascade), and the last column shows the measured cross section.}
    \begin{tabular}{ c@{\hskip 12pt} c @{\hskip 12pt}c@{\hskip 12pt} c@{\hskip 12pt} c}\hline\hline
        \begin{tabular}{@{}c@{}} \rule{0pt}{0.3cm}Initial Beam Energy\\ (MeV/nucleon)\end{tabular} &  $E_{eff}$ (MeV)  & \begin{tabular}{@{}c@{}} \rule{0pt}{0.3cm} Integrated Beam \\ Particles \end{tabular}& \begin{tabular}{@{}c@{}} \rule{0pt}{0.3cm}Detection\\ Efficiency (\%)\end{tabular} & $\sigma$ (mb) \\\hline
        \rule{0pt}{0.35cm}    3.7 & $2.99^{+0.03}_{-0.06}$ & (2.15 $\pm$ 0.11)$\times 10^{11}$& 51.6 $\pm$ 5.6 & 1.63 $\pm$ 0.40\\
        \rule{0pt}{0.3cm}    3.4 & $2.68^{+0.03}_{-0.06}$ & (2.05 $\pm$ 0.10)$\times 10^{11}$& 51.3 $\pm$ 5.7 & 0.72 $\pm$ 0.16\\
        \rule{0pt}{0.3cm}    3.1 & $2.38^{+0.02}_{-0.09}$ & (2.07 $\pm$ 0.10)$\times 10^{11}$& 52.6 $\pm$ 6.0 & 0.23 $\pm$ 0.04\\\hline\hline
    \end{tabular}
    \label{cs_table}
\end{table*}
\par The uncertainties in the presented cross section include: 5\% for the beam-charge accumulation, 5\% for the areal target density attributed to the measurement of the gas cell pressure, statistical uncertainty varying between 1\% and 4\%, with the latter value corresponding to the smaller energy. The overall uncertainty was between 19\% and 24\% as shown in Table~\ref{cs_table}. The largest contribution to this uncertainty comes from the ratio of the yield of the measurement over the detector's efficiency, as described in Sec.~\ref{efficiency_yield}. The uncertainty in the effective energy is mainly attributed to the energy straggling of the \nuc82Kr beam when passing through the Mo foil (2-3\%) and hydrogen gas (1\%). The asymmetric errors in the effective energy result from the asymmetric energy straggling distribution \cite{srim}. The uncertainty from the theoretical cross section input in the calculation of the effective energy as described in Ref.~\cite{cauldrons} does not exceed 15 keV ($<$0.5\%).

\par In Fig.~\ref{fig:cs} the measured cross section is compared to theoretical calculations using the \textsc{non-smoker} \cite{nonsmoker} code shown with the blue solid line and the \textsc{talys} 1.96 \cite{talys} code shown in the orange band. The orange \textsc{talys} band includes all the available NLD and E1 \g SF models in the code (so called ``ldmodel" and ``strength" options). The comparison indicates that standard statistical model calculations with the default models tend to overproduce the cross section of the \nuc82Kr(p,\g)\nuc83Rb reaction relative to the experimentally measured values. The deviations of the reported data from the values predicted by the \textsc{non-smoker} code vary between 23\% and 47\% with the latter corresponding to the lowest beam energy. 

\par A similar behavior in the comparison between measured cross-section data and standard statistical model calculations has been observed in the recently published data from Lotay \textit{et al.} \cite{Lotay_2021} for the (p,\g) reaction on the neighboring \nuc83Rb nucleus, as well as on the data published by Gy\"urky \textit{et al.} \cite{gyurky} on various proton-rich Sr isotopes, using different experimental setups and techniques. In both cases the reported experimental cross section tends to be smaller than the values predicted by the HF theory.
\par This significant overproduction of the cross section by theoretical calculations motivated further investigation, as described in the following section.

\section{\label{sec:theoretical}Theoretical Analysis}

The comparison of the measured cross section with the theoretical calculations from \textsc{talys} shown in Fig.~\ref{fig:cs} indicates an overestimation of the evaluated cross section. However, these \textsc{talys} calculations were performed varying only the NLD and \g SF models for the default parameters chosen by the code. 

\par As discussed in Section \ref{efficiency_yield}, a choice of parameters for modeling the NLD and \g SF of the \nuc83Rb nucleus was made. These parameters were chosen to replicate the experimental SoS spectra obtained by the de-excitation of the \nuc83Rb CN decay and, therefore, should provide a better description of the cross section. In Table \ref{par_table} a few of the chosen NLD and \g SF parameter combinations are provided, along with the default CT and BSFG parameters \cite{E&B2009}, for comparison. 
\begin{table}[ht!]
    \centering
    \caption{Parameters for modeling the NLD and \g SF of the \nuc83Rb nucleus. The default parameters in the first two rows are shown in Fig.~\ref{fig:chi2}(a) with a red and blue line. The rest of the parameters were chosen for this analysis, as described in Section \ref{efficiency_yield}, and form the green band shown in Fig.~\ref{fig:chi2}(a). See text for details on parameters.}

    \begin{tabular}{l c| c l |c l}\hline\hline
        \multicolumn{2}{c}{NLD Model}  &\multicolumn{2}{|c|}{\begin{tabular}{@{}c@{}} \rule{0pt}{0.3cm}NLD Model\\ Details\end{tabular}} & \multicolumn{2}{c}{Upbend in \g SF}\\ \hline\hline
        \multicolumn{2}{c|}{CT default} & \begin{tabular}{@{}l@{}} \rule{0pt}{0.3cm} T = 0.824 \\ $E_0$ = -1.16 \end{tabular} &{\cite{E&B2009}} &\multicolumn{2}{c}{No}\\[.4cm]
        \multicolumn{2}{c|}{BSFG default} & \begin{tabular}{@{}l@{}} \rule{0pt}{0.3cm} $\alpha$ = 10.17 \\ $\Delta$ = - 0.54 \end{tabular} &{\cite{E&B2009}} &\multicolumn{2}{c}{No}\\\hline
        1.& CT & \begin{tabular}{@{}c@{}} \rule{0pt}{0.3cm} T = 0.824 \\ $E_0$ = -2.2 \end{tabular} &&\multicolumn{2}{c}{No}\\[.4cm]
        2.& CT & \begin{tabular}{@{}c@{}} \rule{0pt}{0.3cm} T = 0.861 \\ $E_0$ = -3.34\end{tabular}& {\cite{hoffman2004}} &\multicolumn{2}{c}{No}\\[.4cm]
        3.& BSFG &\begin{tabular}{@{}c@{}} \rule{0pt}{0.3cm} $\alpha$ = 10.17 \\ $\Delta$ = -1.6 \end{tabular}& &\multicolumn{2}{c}{No}\\[.4cm]
        4.& BSFG &\begin{tabular}{@{}c@{}} \rule{0pt}{0.3cm} $\alpha$ = 10.17 \\ $\Delta$ = -0.54 \end{tabular} && \begin{tabular}{@{}c@{}} a = 1.5  \\ c = $8.7\times 10^{-8}$ \end{tabular} &  \cite{mo_upbend} \\[.4cm]
        5.& BSFG & \begin{tabular}{@{}c@{}} \rule{0pt}{0.3cm} $\alpha$ = 10.17 \\ $\Delta$ = -0.54 \end{tabular}&& \begin{tabular}{@{}c@{}} \rule{0pt}{0.3cm} a = 1.0 \\ c = $1.0\times 10^{-7}$ \end{tabular}&\\ \hline\hline
    \end{tabular}
    \label{par_table}
\end{table}
\par The total NLD as a function of the excitation energy, $\rho(E_x)$, as described in the CT model \cite{ctm,ctm2} is 
\begin{equation}\label{eq:ct}
    \rho_\text{CT}(E_x)=\frac{1}{T}\text{exp}\left(\frac{E_x-E_0}{T}\right)
\end{equation} 
where the temperature, $T$, and $E_0$ are free parameters.
The total NLD of the BSFG model \cite{bsfg,bsfg2} is 
\begin{equation}\label{eq:bsfg}
    \rho_\text{BSFG}(E_x)=\frac{1}{12\sqrt{2}\sigma\alpha^{1/4}}\frac{\text{exp}\left[2\sqrt{\alpha(E_x-\Delta)}\right]}{(E_x-\Delta)^{5/4}}
\end{equation}
where $\sigma$ is the spin cut-off parameter, and $\alpha$ and $\Delta$ are free parameters that can be altered.
Regarding the \g SF, for all of the combinations listed in Table \ref{par_table}, 
the E1 and M1 strength parameters were obtained through Ref.~\cite{ripl}, and the E2 strength through Ref.~\cite{talys_manual}. In some occasions shown in Table \ref{par_table}, a low-energy upbend was implemented:
\begin{equation}\label{eq:upbend}
    f_\text{upbend} = c \, \text{exp}[-a E_\gamma]
\end{equation}
where $c$ and $a$ are a normalization and an energy-dependent factor for the low-energy upbend of the \g SF \cite{talys_manual}.
\par It is interesting to note that those combinations of NLD and \g SF which describe the SoS, TAS, and multiplicity spectra (green bands in Fig.~\ref{fig:chi2}) correspond to the upper range of the calculated \nuc82Kr(p,\g )\nuc83Rb\ cross sections (see orange band in Fig.~\ref{fig:cs}). As the new experimental data are located at the lower end of the orange band in Fig.~\ref{fig:cs}, there seems to be some tension between a reasonable description of SoS, TAS, and multiplicity spectra on the one hand and the (p,\g ) cross sections on the other hand. Obviously, this finding calls for a more detailed theoretical analysis.

\par In a schematic notation, the cross section of the \nuc82Kr(p,\g )\nuc83Rb reaction in the statistical model is given by
\begin{equation}
\sigma(p,\gamma) \sim \frac{T_{p,0} T_\gamma}{\sum_i T_i} w_{\rm{WFCF}} = T_{p,0}  b_\gamma  w_{\rm{WFCF}}
\label{eq:StM}
\end{equation}
with the transmissions $T_X$ into the channel $X$ ($X$ = p, n, \g , $\alpha$, etc.), the \g -branching $b_\gamma = T_\gamma / \sum_i T_i$, and the width-fluctuation correction factor $w_{\rm{WFCF}}$. At the low energies under experimental study, the only open channels are the proton and the \g\ channel. The neutron channel opens slightly above 5 MeV. The $\alpha$ channel remains negligible at all energies under study because of the higher Coulomb barrier. The relevance of the different exit channels and the resulting sensitivities on the chosen input parameters of the statistical model will be discussed in the following paragraphs.

\begin{figure}[ht!]
    \includegraphics[width=0.48\textwidth]{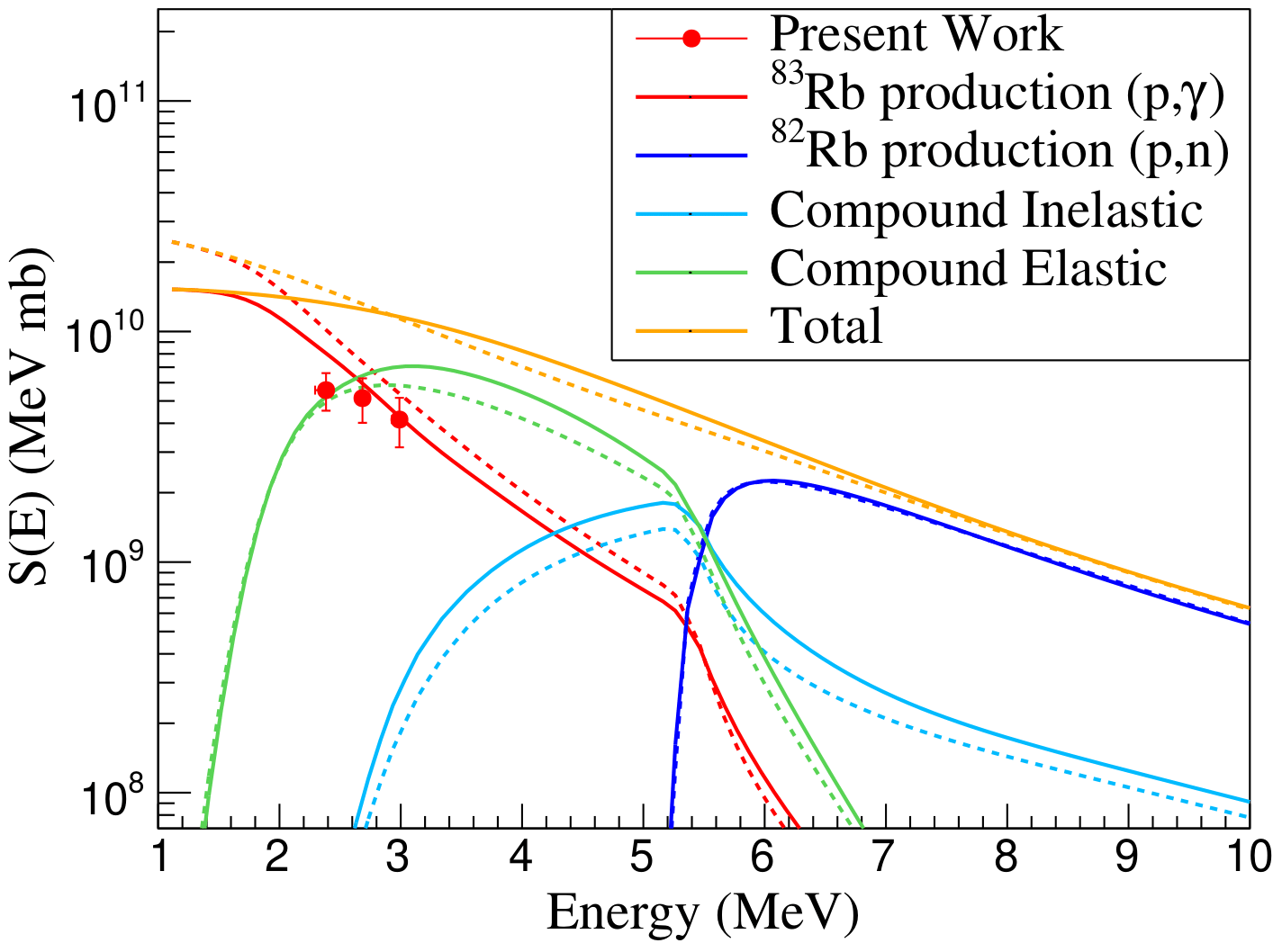}
    \caption{Decomposition plot of the S-factor for proton capture on \nuc82Kr. The dashed lines correspond to a standard \textsc{talys} calculation. The solid lines use an optimized set of parameters. See text for more details on the optimized parameters. The \nuc79Br production by the \nuc82Kr(p,$\alpha$)\nuc79Br reaction is below the scale of the figure.}
    \label{fig:decomp}
\end{figure}

\par In such an analysis, it is useful to investigate, not just the channel of interest, but also additional reaction channels, to evaluate the competition between them. This also provides an intuitive understanding of the sensitivities of the calculated cross sections on the different ingredients of the statistical model. Figure \ref{fig:decomp} shows a decomposition of the total reaction cross section $\sigma_{\rm{tot}}$ into the different exit channels. For better readability, the cross sections are converted to astrophysical S-factors. The combination of NLD and \g SF used in this plot is listed as No. 5 in Table \ref{par_table}. The dashed lines correspond to a standard calculation; the full lines use an optimized set of parameters (as discussed below). Already close above the opening of the neutron channel (slightly above 5 MeV), the neutron channel dominates, leading to a neutron branching $b_n \approx 1$ and $\sigma(p,n) \approx \sigma_{\rm{tot}}$. In contrast, at the energies under experimental study, the situation is more complicated because the dominating channels of (p,\g ) proton capture and (p,p) compound-elastic scattering show comparable strengths. Under these circumstances, Eq.~(\ref{eq:StM}) simplifies to
\begin{equation}
\sigma(p,\gamma) \sim \frac{T_{p,0} T_\gamma}{T_{p,0} + T_\gamma} w_{\rm{WFCF}}
\label{eq:StM2}
\end{equation}
As the NLD and \g SF are constrained by the SoS, TAS, and multiplicity spectra in Fig.~\ref{fig:chi2}, the \g\ transmission $T_\gamma$ is essentially fixed in Eq.~(\ref{eq:StM2}), and consequently the calculated (p,\g ) cross section remains sensitive only to the proton transmission in the entrance channel $T_{p,0}$, i.e., the proton optical model potential (POMP), and the width-fluctuation correction (WFC) factor $w_{\rm{WFCF}}$.

\par As the total reaction cross section $\sigma_{\rm{tot}}$ depends only on the POMP, it is a simple task to determine the influence of the POMP. The different global POMPs in \textsc{talys} show only minor variations for the resulting $\sigma_{\rm{tot}}$ and thus also on the (p,\g ) capture cross section. Nevertheless, the energy dependence of the so-called ``jlm-type" potentials (based on the work of Jeukenne, Lejeunne, and Mahaux \cite{Jeu74,Jeu76,Jeu77a,Jeu77b} with later modifications by Bauge {\it {et al.}} \cite{Bau98,Bau01}) shows slightly lower cross sections at the lowest energies, leading to a better agreement with the new experimental data.

\par
\begin{figure}[ht!]
    \centering
    \includegraphics[width=0.48\textwidth]{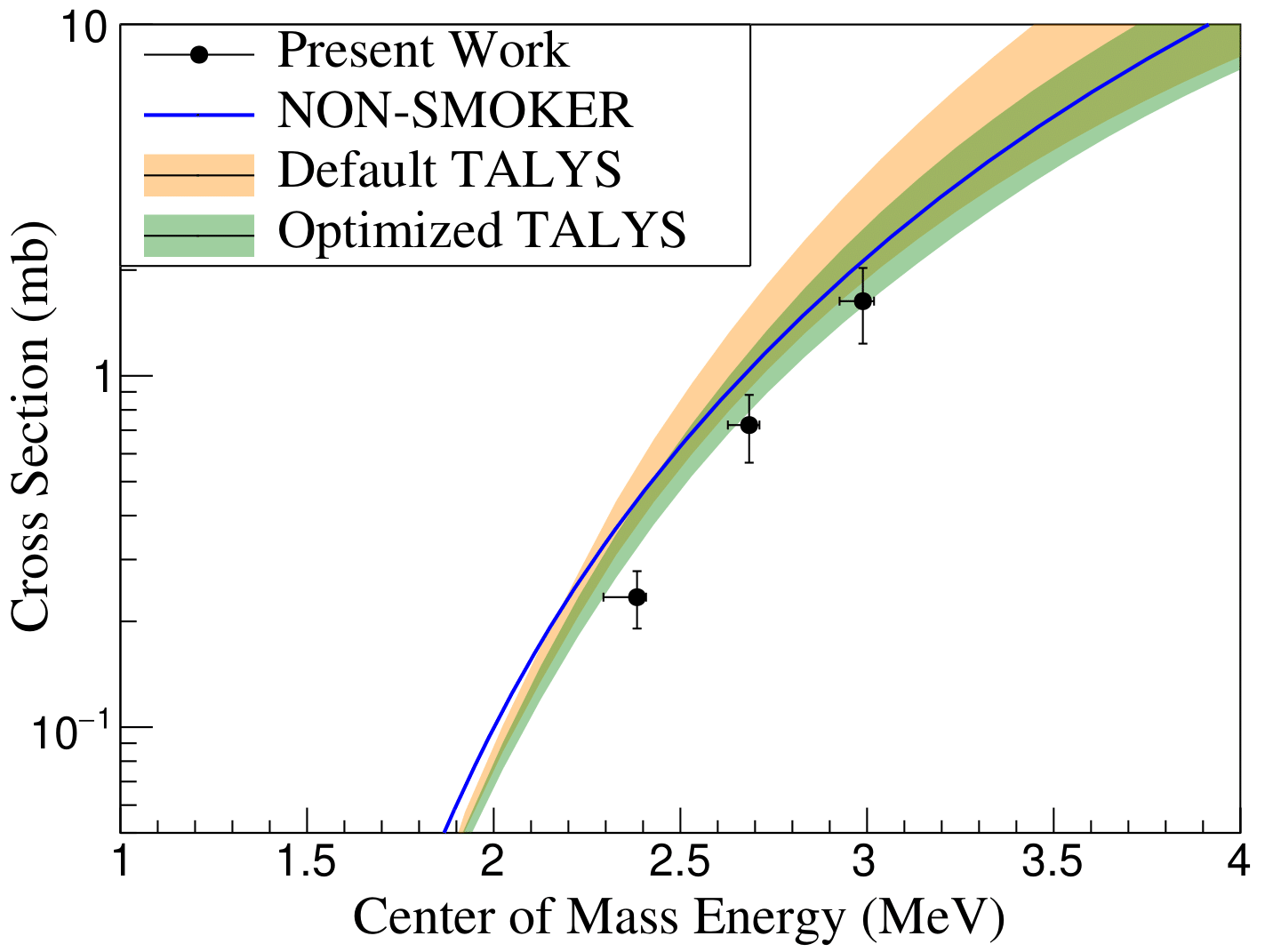}
    \caption{Same as Fig.~\ref{fig:cs} with the addition of the green band, which corresponds to optimized \textsc{talys} calculations using the parameters for modeling the NLD and \g SF of the \nuc83Rb nucleus from Table~\ref{par_table}, as well as a ``jlm-type" potential and a strong WFC, as described in Section~\ref{sec:theoretical}.}
    \label{fig:cs2}
\end{figure}

 The WFC takes into account that there are correlations between the incident and outgoing wave functions. These correlations typically enhance the compound-elastic channel and reduce the cross sections of the reaction channels. The WFC becomes most pronounced at low energies with only few open channels, whereas at higher energies and many open channels the relevance of the WFC becomes negligible. By default, \textsc{talys} applies a WFC using the formalism of Moldauer (so-called ``widthmode 1") \cite{Mol76,Mol80}. A much stronger WFC is obtained for the approach of Hofmann, Richert, Tepel, and Weidenm\"uller (HRTW approach, ``widthmode 2") \cite{Tep74,Hof75,Hof80}, leading to significantly lower calculated (p,\g ) cross sections, especially at low energies. Thus, the WFC using the simple HRTW approach shows much better agreement with the new experimental data, as can be seen from the green band in Fig.~\ref{fig:cs2}.

\par The above study shows that a consistent description of the \nuc82Kr(p,\g )\nuc83Rb\ cross section and the SoS, TAS, and multiplicity spectra can be obtained using a careful choice of parameters in the statistical model. The following parameters have finally been adopted: The POMP was taken from Jeukenne, Lejeune, and Mahaux with the Bauge modification (so-called ``jlmmode 3" in \textsc{talys}); the $\alpha$ optical model potential was kept as default because it has practically no influence on the (p,\g ) cross sections under study; the width fluctuation correction is based on the HRTW approach; good combinations of the NLD and the \g SF are listed in Table \ref{par_table}. The green band in Fig.~\ref{fig:cs2} was created using this choice of parameters. The curves in Fig.~\ref{fig:decomp} are obtained from combination No.~5 in Table \ref{par_table}.

\par As a final remark, the simultaneous analysis of SoS, TAS, and multiplicity spectra provides much stronger constraints on the statistical properties of the produced compound nucleus than a standard analysis of (p,\g ) cross sections. The adopted technique can constrain the product of two of the most important ingredients to the statistical HF model, the NLD and \g SF. \break

\section{Summary \& Conclusions}

\par The total cross section of the \nuc82Kr(p,\g)\nuc83Rb reaction was measured in inverse kinematics using a stable \nuc82Kr beam at effective energies between $\approx$ 2.4 and 3.0 MeV. The obtained spectra were analyzed using the \g-summing technique. The large width of the sum peak due to the Doppler effect and energy straggling through the gas-cell target required the detection efficiency of this measurement to be calculated as a function of the contribution of each possible excitation energy of the \nuc83Rb CN. For this reason, a new analysis technique developed in  Ref.~\cite{Alicia} was applied. The extracted cross section was compared to standard HF statistical model calculations using the \textsc{non-smoker} and \textsc{talys} codes for default inputs of NLD and \g SF. The comparison indicates that standard statistical model calculations tend to overproduce the cross section of the \nuc82Kr(p,\g)\nuc83Rb reaction relative to the experimentally measured values. A similar behavior has been observed for neighboring nuclei by Refs.~\cite{Lotay_2021,gyurky}, thus motivating the authors' further theoretical investigation on the choice of parameters in the statistical model. Choosing a special width fluctuation correction, a consistent description of the \nuc82Kr(p,\g )\nuc83Rb\ cross section and the experimental spectra was obtained. The presented analysis technique can provide stronger constraints for the choice of the parameters for the statistical model calculations than a simple comparison to the cross section measurement.

\section*{ACKNOWLEDGMENTS}
\par The authors would like to acknowledge the support of the ReA3 accelerator team with optimizing the beam delivery and setup.
\par This work was supported by the National Science Foundation under grants No. PHY 1613188 (Hope College), No. PHY 1102511 (NSCL), No. PHY 1913554 (Windows on the Universe: Nuclear Astrophysics at the NSCL), No. PHY 2209429 (Windows on the Universe: Nuclear Astrophysics at FRIB), No. PHY 1430152 (Joint Institute for Nuclear Astrophysics) and by NKFIH (K134197). 
\par This material is based upon work supported by the U.S. Department of Energy, National Nuclear Security Administration through grant No. DOE-DENA0003906, award No. DE-NA0003180 (Nuclear Science and Security Consortium) as well as grant No. DE-FG0296ER40963 and No. DE-SC0020451 from the Department of Energy, Office of Science, Office of Nuclear Physics.
\par \hfill 

\bibliography{mybib}

\end{document}